%Paper: gr-qc/9407021
%From: Steve Poletti <spoletti@physics.adelaide.edu.au>
%Date: Sat, 16 Jul 1994 10:58:32 +0930 (CST)
%Date (revised): Fri, 9 Dec 1994 15:44:34 +1030 (CST)

%There is one figure which is available on request from the authors

\input phyzzx \def\exs{{}}
%% If use A4 sized paper - i.e. you don't live in North America - you may
%% decrease the length of the paper by up to 3 pages (depending on your
%% version of phyzzx.tex) by uncommenting the following line:
%\referenceminspace=36pt\vsize=10true in\voffset=-5true mm\def\exs{\vfil\eject}

\overfullrule=0pt\hsize=6.5truein
\rightline{\vbox{\halign{#\hfil\cr ADP-94-2/M22\cr gr-qc/9407021\cr}}}
\bigskip\bigskip\baselineskip=24pt
\title{\seventeenbf Global properties of static spherically symmetric
\break charged dilaton spacetimes with a Liouville potential}
\baselineskip=18pt
\vfill
\author{S.J. Poletti\foot{E-mail: spoletti@physics.adelaide.edu.au} and
D.L. Wiltshire\foot{E-mail: dlw@physics.adelaide.edu.au} }
\address{Department of Physics and Mathematical Physics, University of
Adelaide,\break Adelaide, S.A. 5005, Australia.}
\vfill

%%%%%%%%%%%%%%%%%%%%%%%%%%%%%%%%%%%%%%%%%%%%%%%%%%%%%%%%%%%%%%%%%%%%%%%%%%%%%%%
\font\sixrm=cmr6 \font\sixi=cmmi6 \font\sixsy=cmsy6 \font\sixbf=cmbx6
\font\eightrm=cmr8 \font\eighti=cmmi8 \font\eightsy=cmsy8 \font\eightbf=cmbx8
\font\eighttt=cmtt8 \font\eightit=cmti8 \font\eightsl=cmsl8

\def\tenpoint{\def\rm{\fam0\tenrm} \textfont0=\tenrm \scriptfont0=\sevenrm
\scriptscriptfont0=\fiverm \textfont1=\teni \scriptfont1=\seveni
\scriptscriptfont1=\fivei \textfont2=\tensy \scriptfont2=\sevensy
\scriptscriptfont2=\fivesy \textfont3=\tenex \scriptfont3=\tenex
\scriptscriptfont3=\tenex \textfont\itfam=\tenit \def\it{\fam\itfam\tenit}
\textfont\slfam=\tensl \def\sl{\fam\slfam\tensl} \textfont\ttfam=\tentt
\def\tt{\fam\ttfam\tentt} \textfont\bffam=\tenbf \scriptfont\bffam=\sevenbf
\scriptscriptfont\bffam=\fivebf \def\bf{\fam\bffam\tenbf}
\setbox\strutbox=\hbox{\vrule height8.5pt depth3.5pt width0pt}
\let\sc=\eightrm \let\big=\tenbig \rm} \def\eightpoint{\def\rm{\fam0\eightrm}
\textfont0=\eightrm \scriptfont0=\sixrm \scriptscriptfont0=\fiverm
\textfont1=\eighti \scriptfont1=\sixi \scriptscriptfont1=\fivei
\textfont2=\eightsy \scriptfont2=\sixsy \scriptscriptfont2=\fivesy
\textfont3=\tenex \scriptfont3=\tenex \scriptscriptfont3=\tenex
\textfont\itfam=\eightit \def\it{\fam\itfam\eightit} \textfont\slfam=\eightsl
\def\sl{\fam\slfam\eightsl} \textfont\ttfam=\eighttt
\def\tt{\fam\ttfam\eighttt} \textfont\bffam=\eightbf \scriptfont\bffam=\sixbf
\scriptscriptfont\bffam=\fivebf \def\bf{\fam\bffam\eightbf}
\setbox\strutbox=\hbox{\vrule height7pt depth2pt width0pt} \let\sc=\sixrm
\let\big=\eightbig \rm} \def\br{\hfil\break} \def\rarr{\rightarrow}
\def\scrscr{\scriptscriptstyle}  
\def\begincaption{\medskip\openup-1\jot\eightpoint} \def\scr{\scriptstyle}
\def\endcaption{\tenpoint\openup1\jot\leftskip=0pt\rightskip=0pt}
\def\caption#1#2{\message{#1}\begincaption\leftskip=15true mm\rightskip=15true
mm\vbox{\halign{\vtop{\parindent=0pt\parskip=0pt\strut##\strut}\cr{\bf#1}\quad
#2\cr}}\endcaption} \def\dsp{\displaystyle} \def\e{{\rm e}}
\def\sqr#1#2#3{{\vbox{\hrule height.#2pt \hbox{\vrule width.#2pt
height#1pt\kern#1pt\vrule width.#2pt}\hrule height.#2pt}\hbox{\hskip.#3em}}}
\def\Dal{\,{\mathchoice\sqr64{15}\sqr64{15}\sqr431\sqr331}} \def\et{\eta}
\def\goesas{\mathop{\sim}\limits}\def\Rarr{\Rightarrow}\def\w#1{\;\hbox{#1}\;}
\def\Z#1{_{\lower2pt\hbox{$\scr#1$}}} \def\X#1{_{\lower2pt\hbox{$\scrscr#1$}}}
\def\ns#1{_{\hbox{\sevenrm #1}}} \def\const{\hbox{const.}}\def\al{\alpha}
\def\be{\beta}\def\ga{\gamma}\def\ep{\epsilon}
\def\ee{\varepsilon}\def\ka{\kappa}\def\th{\theta}\def\ph{\phi}
\def\ch{\chi}\def\la{\lambda}\def\rh{\rho}\def\pt{\partial}
\def\OM{\Omega}\def\dd{{\rm d}}\def\LA{\Lambda}
\def\ze{\zeta}\def\PL#1{Phys. Lett.\ {\bf#1}}\def\CMP#1{Commun.\ Math.\ Phys.\
{\bf#1}} \def\AP#1#2{Ann.\ Phys.\ (#1)
{\bf#2}} \def\PR#1{Phys.\ Rev.\ {\bf#1}}\def\CQG#1{Class.\ Quantum Grav.\
{\bf#1}}\def\NP#1{Nucl.\ Phys.\ {\bf#1}} \def\IJMP#1{Int.\ J. Mod.\ Phys.\ {\bf
#1}} \def\MPL#1{Mod.\ Phys.\ Lett.\ {\bf #1}} \def\RN{Reissner-Nordstr\"om}
%%%%%%%%%%%%%%%%%%%%%%%%%%%%%%%%%%%%%%%%%%%%%%%%%%%%%%%%%%%%%%%%%%%%%%%%%%%%%%%

\abstract

We derive the global properties of static spherically symmetric solutions to
the Einstein-Maxwell-dilaton system in the presence of an arbitrary exponential
dilaton potential. We show that -- with the exception of a pure cosmological
constant `potential' -- no asymptotically flat, asymptotically de Sitter or
asymptotically anti-de Sitter solutions exist in these models.

\vfill\endpage \baselineskip=16pt
\REF\GM{G.W. Gibbons and K. Maeda, \NP{B298} (1988) 741.}

\REF\GHS{D. Garfinkle, G.T. Horowitz and A. Strominger, \PR{D43} (1991) 3140;
err.\ ibid.\ {\bf 45} (1992) 3888.}

\REF\HL{B. Harms and Y. Leblanc, \PR{D46} (1992) 2334.}%hep-th/9205021

\REF\HW{C.F.E. Holzhey and F. Wilczek, \NP{B380} (1992) 447.}%hep-th/9202014

\REF\DP{T. Damour and A.M. Polyakov, \NP{B423} (1994) 532.}%hep-th/9401069.

\REF\GH{R. Gregory and J.A. Harvey 1993, \PR{D47} (1993) 2411.}%hep-th/9209070

\REF\HH{J.H. Horne and G.T. Horowitz 1993, \NP{B399} (1993) 169.}
%hep-th/9210012

\REF\DIN{J.P. Derendinger, L.E. Ib\'a\~nez and H.P. Nilles, \PL{155B} (1985)
65; \br M. Dine, R. Rohm, N. Seiberg and E. Witten, \PL{156B} (1985) 55.}
% For some more recent work see, e.g.,
% \REF{Casas J.A., Lalak, Z., Mu\~noz C. and Ross, G.G. 1990 Heirachical
% supersymmetry breaking and dynamical determination of compactification
% parameters by non-perturbative effects \NP{B347}, 243.\br
% Font A., Ib\'a\~nez L.E., L\"ust D. and Quevedo, F. 1990 Supersymmetry
% breaking from duality invariant gaugino condensation \PL{245B}, 401.\br
% Cveti\v{c} M., Font A., Ib\'a\~nez L.E., L\"ust D. and Quevedo, F. 1991
% Target-space duality, supersymmetry breaking and the stability of classical
% string vacua \NP{B361}, 194-232}

\REF\cosm{S. Kalara and K.A. Olive, \PL{218B} (1989) 148.\br
M.C. Bento, O. Bertolami and P.M. S\'a, \PL{262B} (1991) 11; \MPL{A7} (1992),
911.\br
M. Gasperini and G. Veneziano, \PL{277B} (1992) 256.\br %hep-th/9112044
A.A. Tseytlin and C. Vafa, \NP{B372} (1992) 443.\br
A.A. Tseytlin, \CQG{9} (1992) 979; \IJMP{\bf D1} (1992) 223.\br %hep-th/9203033
J. Garc\'\i a-Bellido and M. Quir\'os, \NP{B368} (1992) 463; \NP{B385}
(1992) 558.\br %hep-th/9204079
N. Kaloper and K.A. Olive, Astropart.\ Phys.\ {\bf1} (1993) 185.}
%``Dilatons in string cosmology''

\REF\MWa{S. Mignemi and D.L. Wiltshire, \CQG{6} (1989) 987.}

\REF\Wi{D.L. Wiltshire, \PR{D44} (1991) 1100.}

\REF\MWb{S. Mignemi and D.L. Wiltshire, \PR{D46} (1992) 1475.}%hep-th/9202031

\REF\BC{F. Belgiorno and A.S. Cattaneo, ``Black holes and cosmological constant
in bosonic string theory: some remarks'', Preprint IFUM 450/FT,
hep-th/9309156.}

\REF\Rama{S.K. Rama, ``Cosmological constant in low energy d=4 string leads to
naked singularity'', Preprint TCD-1-94, hep-th/9402009.}

\REF\MS{T. Maki and K. Shiraishi, \CQG{10} (1993) 2171.}

\REF\KT{D. Kastor and J. Traschen, \PR{D47} (1993) 5370.}%hep-th/9212035

\REF\CM{M. Cadoni and S. Mignemi, \PR{D48} (1993) 5536; %hep-th/9305107
\NP{B427} (1994) 669.}%hep-th/9312171

\REF\CT{M. Cveti\v{c} and A.A. Tseytlin, \NP{B416} (1994) 137.}%hep-th/9307123.

\REF\Ok{T. Okai, ``4-dimensional dilaton black holes with cosmological
constant'', Preprint UT-679, hep-th/9406126}

\REF\Ch{J.E. Chase, \CMP{19} (1970) 276.}

\REF\Be{J.D. Bekenstein, \PR{D5} (1972) 1239; 2403.}

\REF\Gary{G.W. Gibbons ``Self-gravitating magnetic monopoles, global monopoles
and black holes'', in J.D. Barrow, A.B. Henriques, M.T.V.T. Lago and M.S.
Longair (eds.), {\it The Physical Universe: Proceedings of the XII Autumn
School, Lisbon, 1990}, (Springer, Berlin, 1991).}

\REF\GW{G.W. Gibbons and D.L. Wiltshire, \AP{N.Y.}{167} (1986) 201; err.\
ibid.\ {\bf 176} (1987) 393.}

\REF\Bu{H. Buchdahl, \PR{115} (1959) 1325.}

\REF\JSD{S.J. Poletti, J. Twamley and D.L. Wiltshire, preprint ADP-94-17/M24,
hep-th/9412076.}

\def\g#1{{\rm g}\Z#1} \def\V{{\cal V}} \def\ddim{$D$-dimensional\ }
\def\cd#1{c\Z#1} \def\dVdph{{\dd\V\over\dd\ph}} \def\Ve{\V\ns{exp}}
\def\Vs{\V\ns{susy}} \def\kp{2\kappa\phi} \def\vph#1{\vphantom{#1}}
\def\lb{\bar\la} \def\Dm#1{D-#1} \def\Db#1{(\Dm#1)} \def\pdd{\Db3\g1-\g0}
\def\pyy{1+\Db3\g1^{\ 2}} \def\gg{\g0^{\ 2}} \def\pvv{\Dm3+\gg}
\def\pvy{1+\g0\g1} \def\pvvB{\left(\pvv\right)} \def\rH{r\X{\cal H}}
\def\cP{{\cal P}} \def\ggg{\g1^{\ 2}} \def\gbb{\left(\g0+\g1\right)}
\def\G#1{{\rm g}\X#1} \def\GG{\G0^{\ 2}} \def\GGG{\G1^{\ 2}}
\def\rhj{\bar\rh\left(1-\bar\rh\right)^{-1}} \def\sign{\hbox{sign}\;}
\def\LAe{\ee\LA}\def\lbe{\ee\lb}\def\Qe{\ee Q^2}\def\epk#1{\ep\Z#1k\Z#1^{\ 2}}
\def\xixi#1{\left(\xi-\xi\X#1\right)} \def\rb{\bar r} \def\rrp{\rb+\rb\Z0}
\def\rrm{\rb-\rb\Z0} \def\rrmp{\left(\rrm\over\rrp\right)} \def\pVV{\Dm3+\GG}
\def\muck{\left\{\rrmp^c-A^2\rrmp^{-c}\right\}}
\def\rra{\left(\rb\over\bar a\rb+1\right)}
\chapter{Introduction}

There has been considerable interest recently in the properties of `stringy'
black holes: classical solutions of tree-level string effective actions, in
which the Einstein action is supplemented by fields such as the axion,
gauge fields, and the dilaton which couples in a
 non-trivial way to the other fields. In particular, dilaton
black holes have been shown to have novel thermodynamic properties [\GM--\HL],
and to behave like elementary particles in some scattering scenarios [\HW].

Unfortunately, much of the work on dilaton black holes to date has involved
models with one serious deficiency: the dilaton has usually been assumed to be
massless. It is widely believed, however, that the dilaton must aquire a mass
through some symmetry breaking mechanism. Indeed, this is necessary in order to
avoid long-range scalar forces which would otherwise arise\foot{A fascinating
discussion of the observational consequences of a very weakly coupled massless
dilaton is given in [\DP].}. Gregory and Harvey
[\GH] and Horne and Horowitz [\HH] have now finally made some investigation of
black hole models which include a mass term -- they have chosen a standard
quadratic potential for the dilaton field\foot{Gregory and Harvey also
considered a potential of the form $\V=2m^2\cosh^2\phi$.}. While a rigorous
proof of the existence of black hole solutions in these models has still to be
given, the arguments of Horne and Horowitz [\HH] are nonetheless compelling.

Ultimately, it would be physically desirable to investigate models of black
holes in effective dilaton gravity theories in four dimensions which involve
a dilaton potential generated by some specific symmetry breaking mechanism,
rather than simply an ad hoc potential, as in the work to date [\GH,\HH]. In
particular, one could consider an (Einstein frame) action such as
$$S=\int\dd^Dx\sqrt{-g}\left\{{{\cal R}\over4\ka^2}-{1\over\Dm2}\,g^{ab}\pt_a
\ph\,\pt_b\ph-\V(\ph)-{1\over4}\exp\left(-4\g0\ka\ph\over\Dm2\right)F_{ab}F^
{ab}\right\},\eqn\action $$
which includes gravity, an abelian gauge field and the dilaton, $\ph$,
with a non-trivial dilaton potential $\V(\ph)$ of the form
$$\V=\Ve+\Vs$$
where
$$\Ve={\LA\over2\ka^2}\exp\left(-4\g1\ka\ph\over\Dm2\right)\,, \eqn\Vexp$$
and
$$\Vs={1\over4\ka^2}\exp\left[-ae^{-\kp}\right]\left\{Ae^{\kp}+B+Ce^{-\kp}
\right\}\,,\eqn\Vsusy$$
Here $\g0$, $\g1$, $\LA$, $a$, $A$, $B$ and $C$ are constants and $\ka^2=4\pi
G$ is the \ddim gravitational constant. Eqns.\ \action--\Vexp\ are somewhat
more general than is demanded by string theory. However, if we set $\g0=1$ we
obtain the standard tree-level coupling between the dilaton and the
electromagnetic field, while setting $\g1=-1$, $\LA=\left(D_{\hbox{\sevenrm
crit}}-D_{\hbox{\sevenrm eff}}\right)/(3\al')$, in the Liouville-type term
\Vexp\ yields the case of a potential corresponding to a central charge
deficit. The term \Vsusy, on the other hand, is the type of potential which
arises in four dimensions from supersymmetry breaking via gaugino condensation
in the hidden sector of the string theory\foot{The particular potential given
in \Vsusy\ is relevant for one gaugino condensation.} [\DIN]. Potentials of
the form \Vexp\ and \Vsusy\ have been widely studied in string cosmologies
[\cosm], but to date the only investigations of static spherically
symmetric solutions involving such terms have been restricted to the case of
uncharged solutions [\MWa--\Rama]. Maki and Shiraishi [\MS] have recently
derived non-static Kastor-Traschen type [\KT] cosmological multi black hole
solutions for the action \action--\Vexp. However, such solutions were only
obtained for certain special values of the constants $\g0$, $\g1$ and of the
time-dependent coupling in the dilaton cosmological scale factor.

In many respects the action \action\ is still over-simplified because it
neglects the possible contribution of additional scalar fields, such as the
string moduli which correspond to the extra dimensions of spacetime after
dimensional reduction. Static spherically symmetric solutions involving
both moduli and a dilaton have been discussed recently by Cadoni and Mignemi
[\CM], and by Cveti\v{c} and Tseytlin [\CT], but in the absence of a potential.
The introduction of a potential greatly complicates the situation, however, as
is well demonstrated by the case of the quadratic potential, where a complete
integration proved impossible even numerically [\HH].

Given the inherent difficulties involved in studies of models with non-trivial
potentials, the present paper is intended only as a first step: we will not
study the problem posed by eqns.\ \action--\Vsusy\ in full, but will restrict
ourselves solely to the case of a Liouville-type potential $\V=\Ve$. It is our
hope that a further refinement of the approach discussed here can be applied
to the more difficult case when a supersymmetry-breaking potential of the type
\Vsusy\ is also included. We are of course most interested in the case $D=4$,
but will leave $D$ arbitrary, (with the only requirement that $D>2$), as this
does not involve many extra complications. Furthermore, we will also leave the
dilaton coupling to the electromagnetic field arbitrary, rather than
immediately specialising to the string case ($\g0=1$).

We will show in the particular case of an exponential potential
\Vexp\ that no static spherically symmetric asymptotically flat charged black
hole solutions exist. Furthermore, no static spherically symmetric
asymptotically de Sitter (or anti-de Sitter) solutions exist either, except in
the special case $\g1=0$ when the potential is simply a cosmological constant.
(The $\g1=0$ model has recently been studied by Okai [\Ok].)
Our result concerning the potential \Vexp\ is of course more in the line with
the intuition provided by the scalar no hair theorems [\Ch--\Gary], rather than
with the dilaton black hole solutions [\GM] which avoid the no hair theorems
through the coupling between the dilaton and electromagnetic fields, with the
result that the dilaton scalar charge depends on the other charges of theory
[\GW] rather than being an independent ``hair''.
\chapter{The dynamical system}

We shall use the same approach here as has been used by S. Mignemi and one of
us [\MWa--\MWb] to study uncharged static spherically symmetric solutions in
models of gravity involving a scalar field with non-trivial potentials. Such
an analysis is useful for deriving ``no hair'' results in circumstances in
which some assumptions used in the standard no hair proofs do not apply. We
recall, for example, that Bekenstein's proof [\Be] of the scalar no hair
theorem for static black holes can be easily generalised to any convex
potential [\Gary], (i.e., for any $\V(\ph)$ for which ${\dd^2\V\over\dd\phi^2}
\ge0$ for all $\phi$), by simply multiplying the appropriate scalar field
equation by $\dVdph$, rather than by $\ph$, before carrying out the appropriate
integration. In [\MWa], however, we derived the equivalent of a no hair result
for the potential \Vexp\foot{In [\MWa] a particular $\g1$ appropriate to
Kaluza-Klein theories with internal spaces of non-zero curvature was used, but
the arbitrary $\g1$ case was included in [\MWb].} without any restriction on
the sign of $\LA$. (For $\LA>0$ ($\LA<0$) the potential \Vexp\ is convex
(concave).)

In [\Wi] and [\MWb] the approach of [\MWa] was extended to more general
potentials. Although the most general potential we have considered is an
arbitrary finite sum of exponential terms
$$\V(\ph)={-1\over4\ka^2}\;\sum_{i=1}^s\la_i\exp\left(-4\g i\ka\ph\over\Dm2
\right),\eqn\Vexpsum$$
there are some aspects of the analysis of [\MWb] which would appear to apply to
arbitrary potentials. In particular, one may conjecture that:

\noindent (i)\ If $\V$ is non-zero then asymptotically (anti)-de Sitter
solutions exist if and only if
$$\exists\;\ph\Z0\ \w{such that}\ \left.\dVdph\right|_{\ph=\ph\X0}=0\ \w{and}\
\V(\ph\Z0)\ne0.\eqn\conddS$$
The solutions are asymptotically de Sitter (anti-de Sitter) for $\V(\ph\Z0)>0$
($\V(\ph\Z0)<0$).

\noindent (ii)\ If $\V$ is non-zero then asymptotically flat solutions exist
if and only if
$$\exists\;\ph\Z0\ \w{such that}\ \left.\dVdph\right|_{\ph=\ph\X0}=0\ \w{and}\
\V(\ph\Z0)=0.\eqn\condaf$$
The fact that trivial solutions exist under both these circumstances is pretty
obvious: if \conddS\ holds then the Schwarzschild-(anti)-de Sitter solutions
with constant dilaton, $\ph=\ph\Z0$, are solutions; while if \condaf\ is
satisfied then the Schwarzschild solution with constant dilaton, $\ph=\ph\Z0$,
is a solution. Any particular potential may have many such solutions, depending
on the number of different turning points.

It is not difficult to see that any non-trivial solutions with the
appropriate asymptotic form, and with a scalar field which is ``physically
well-behaved'', namely at worst $\ph\goesas\const$ at spatial infinity,
must also satisfy \conddS\ or \condaf\ if all the fields have regular Taylor
expansions at spatial infinity. This can in fact be seen by direct inspection
of the field equations written in terms of a conventional radial coordinate.
In particular, consider coordinates of the type used by Garfinkle, Horowitz
and Strominger [\GHS]:
$$\dd s^2=-e^{2u}\dd t^2+e^{-2u}\dd r^2+R^2\dd\OM^2\Z{\Dm2},\eqn\coorda$$
where $u=u(r)$ and $R=R(r)$. We will henceforth use units in which $\ka=1$. The
field equations obtained from variation of \action\ for a general potential
$\V(\ph)$ are then given by
$$\eqalign{&\Dal\phi=\half\Db2\dVdph-\half\g0\exp\left(-4\g0\ph
\over\Dm2\right)F_{ab}F^{ab},\cr &\pt_a\left[\sqrt{-g}\exp\left(-4\g0\ph
\over\Dm2\right)F^{ab}\right]=0,\cr &{\cal R}_{ab}={4\over\Dm2}\left[\pt
_a\ph\,\pt_b\ph+g_{ab}\V\right]+2\exp\left(-4\g0\ph\over\Dm2\right)
\left[F_{\vph{b}ac}^{\vph{c}}F^{\ c}_b-{\scr 1\over\scr2\Db2}\,g_{ab}^{\vph{c}
}F_{cd}^{\vph{c}}F^{cd}\right],\cr}
\eqn\fieldeq$$
If we choose $\bf F$ to be the field of an isolated electric charge,
$${\bf F}=\exp\left(4\g0\ph\over\Dm2\right){Q\over R^{\Dm2}}\;\dd t\wedge\dd r
\,,$$
then the field
equations with metric ansatz \coorda\ are
$$\eqalign{{1\over R^{\Dm2}}{\dd\ \over\dd r}\left[R^{\Dm2}e^{2u}{\dd\ph\over
\dd r}\right]=\;&\half\Db2\dVdph+\g0\exp\left(4\g0\ph\over\Dm2\right){Q^2
\over R^{2\Db2}}\cr {1\over R}{\dd^2R\over\dd r^2}=\;&-{4\over\Db2^2}\,
\left(\dd\ph\over\dd r\right)^2,\cr {1\over R^{\Dm2}}{\dd\ \over\dd r}\left[e^
{2u}{\dd\ \over\dd r}\left(R^{\Dm2}\right)\right]=\;&\Db2\Db3{1\over R^2}-4
\V-2\exp\left(4\g0\ph\over\Dm2\right){Q^2\over R^{2\Db2}}\,,\cr}
\eqn\eqcoorda$$
together with one further equation which depends on the others by virtue of
the Bianchi identity. Although we have assumed an electric field here, the
solutions for a purely magnetic field are readily obtainable once the
solutions for the system \eqcoorda\ are known since the field equations for the
magnetic case can be obtained from \eqcoorda\ by making the replacement $\g0
\rarr-\g0$ and $Q\rarr P$, where $P$ is the magnetic monopole charge.

If we now make the expansions
$$\eqalign{e^{2u}=\;&{-2\LA r^2\over\Db1\Db2}+u\Z{-1}r
+u\Z0+{u\Z1\over r}+{u\Z2\over r^2}+\dots,\cr
R=\;&r+R\Z0+{R\Z1\over r}+{R\Z2\over r^2}+\dots,\cr
\ph=\;&\ph\Z0+{\ph\Z1\over r}+{\ph\Z2\over r^2}+\dots,\cr}\eqn\expand$$
at spatial infinity, assuming the solutions to be asymptotically (anti)-de
Sitter or asymptotically flat depending on the value of $\LA$, then
substitution of \expand\ in \eqcoorda\ yields the result
$$\left.\dVdph\right|_{\ph=\ph\X0}=0,\qquad\LA=2\V(\ph\Z0),\eqn\asympV$$
from the lowest order terms.
Thus in general it is necessary for the potential to have a turning point at
$\ph=\ph\Z0$ for solutions with asymptotic expansions \expand\ to exist, and
such solutions will be asymptotically flat, de Sitter or anti-de Sitter if
$\V(\ph\Z0)=0$, $\V(\ph\Z0)>0$ or $\V(\ph\Z0)<0$ respectively. Such solutions
are consequently ruled out for the Liouville-type potential \Vexp, except in
the special case of a cosmological constant ($\g1=0$) when $\dVdph=0$
identically.

If we make no assumptions about the existence of regular series expansions
at spatial infinity then the proof of necessity in \conddS\ and \condaf\ is
much less trivial. Indeed, in the context of models of gravity corresponding to
the low-energy limit of string theory one can conceive of instances in which
power series expansions of the form \expand\ would not apply. In particular,
although asymptotics with $\ph\rarr-\infty$ at spatial infinity would be
disastrous in conventional field theories, in the case of string theory all
couplings between the dilaton and matter fields involve powers of $e^{2\ph}$,
so provided the dilaton energy-momentum tensor is well-behaved at spatial
infinity one would expect the weak-coupling limit $\ph\rarr-\infty$ to be
physically admissable. It is under such circumstances that the approach of
[\MWa--\MWb] becomes useful: we reformulate the field equations in terms of
a first order autonomous system of ordinary differential equations. Typically
one finds that the full phase space has various subspaces, one of which
corresponds to the system with no potential and which contains critical points
at the phase space infinity that correspond physically to an asymptotically
flat region. If such critical points are not endpoints for integral curves
which lie outside of the subspace with $\V(\ph)\equiv0$, and if no other
critical points correspond to an asymptotically flat region, then a ``no
hair'' result is obtained. The precise global properties of all solutions of
interest are often readily obtained.

We will now apply this approach to the problem posed by the action
\action--\Vexp. In order to obtain an autonomous system one must use the
radial coordinate of Gibbons and Maeda [\GM], defined by
$$\dd\xi={e^{-2u}\dd r\over R^{\Dm2}}$$
in terms of the previous radial coordinate $r$. One further modification is
necessary for the phase space analysis, namely to replace the $\Db2$-sphere
of the spatial section by a more general $\Db2$-dimensional Einstein space,
so that the full metric is given by
$$\dd s^2=\ee e^{2u}\left[-\dd t^2+R^{2\Db2}\dd\xi^2\right]+R^2g_{ij}\dd x^i\dd
x^j,\eqn\coordb$$
where now $u=u(\xi)$, $R=R(\xi)$,
$${\cal R}_{ij}=\Db3\lb\,g_{ij},\qquad i=1,\dots,\Dm2,$$
and $\ee=\pm1$. Of course, we are primarily interested in the case when $\lb=1$
and $g_{ij}$ is the standard metric for a $\Db2$-sphere. However, the $\lb=0$
surface forms an important boundary in the phase space, with various critical
points lying there, and thus we must leave $\lb$ arbitrary for the analysis.
We have included a factor $\ee=\pm1$ explicitly in \coordb, as this will allow
for the inclusion of critical points both in the region in which the Killing
vector $\pt/\pt t$ is timelike ($\ee=+1$), and the region in which $\pt/\pt t$
is spacelike ($\ee=-1$), in the analysis below.

In terms of the coordinates \coordb\ the field equations are
$$\eqalignno{&\ph''=-\g1\LAe e^{2\ch}+\g0\Qe e^{2\et},&\eqname\junka\cr
&\ze''=\Db3^2\lbe e^{2\ze}-2\LAe e^{2\ch},&\eqname\junkb\cr &u''={-2\over\Dm2}
\LAe e^{2\ch}+2\left(\Dm3\over\Dm2\right)\Qe e^{2\et},&\eqname\junkc\cr
\left(\Dm2\over\Dm3\right)\left[\ze'^2-u'^2\right]-&{4\over\Db2}\ph'^2-
\Db2\Db3\lbe e^{2\ze}+2\LAe e^{2\ch}+\Qe e^{2\et}=0,&\cr
&&\eqname\constraint\cr}$$
where
$$\ze\equiv u+\Db3\ln R,\qquad\et\equiv u+{2\g0\ph\over\Dm2},\qquad \ch\equiv
u+\Db2\ln R-{2\g1\ph\over\Dm2}\,.\eqn\Greekcoord$$
If we now define
$$X=\ze',\qquad Y=\et',\qquad Z=\sqrt{2\over\Dm2}\;e^\et,\qquad V=\ch',\qquad
W=\sqrt{2\over\Dm2}\;e^\ch,\eqn\Latincoord$$
then the constraint equation \constraint\ may be used to eliminate the terms
involving $\lb$ from the field equations \junka--\junkc, yielding the following
first order autonomous dynamical system:
$$\eqalignno{X'&=\Db3\Qe Z^2-\LAe W^2-\Db3\cP&\eqname\autoa\cr
Y'&=\pvvB \Qe Z^2-\left(\pvy\right)\LAe W^2&\eqname\autob\cr
V'&=\left(\Dm3-\g0\g1\right)\Qe Z^2+\left(\ggg-1\right)\LAe W^2-\Db2\cP&
\eqname\autoc\cr
Z'&=YZ&\eqname\autod\cr
W'&=VW&\eqname\autoe\cr}$$
where
$$\eqalign{\cP\equiv{1\over[\pdd]^2}\biggl\{&\left[{\Db1\gg+\Db2^2+2\g0\g
1-\Db3\ggg}\right]X^2\cr
&\!+\left[\pyy\right]Y^2+\pvvB V\left[\Db3V-2\Db2 X\right]\cr
&\!+2(\pvy)Y\left[\Db3V-\Db2X\right]\biggr\}\cr}\eqn\defP$$
and $\g0\ne\Db3\g1$. (We will consider the particular case $\g0=\Db3\g1$
at a later stage).

Our aim is to analyse the phase space for this system of equations. Since the
metric and dilaton field are related to the functions $X$, $Y$, $V$, $Z$ and
$W$ of the 5-dimensional phase space, they are necessarily regular at all
points of the integrals curves apart from critical points. Consequently, in
order to determine the global properties of all solutions -- namely the
structure of their singularities, horizons and asymptotic regions -- it
suffices to study the properties of the solutions at critical points of the
phase space. Further careful analysis is required in order to determine which
critical points are connected to which other ones by integral curves, thus
determining the different possibilities for spacetime structure.

Although the space is 5-dimensional we have some hope of analysing
it due to various symmetries. Equations \autod\ and \autoe\ ensure that
trajectories cannot cross either the $W=0$ or $Z=0$ subspaces, which correspond
physically to $\LA=0$ and $Q=0$ respectively. Thus we may restrict our
attention to $Z\ge0$ and $W\ge0$ without loss of generality.
The hyperboloid defined by $\lb=0$, or equivalently from \constraint
$$\cP-\Qe Z^2-\LAe W^2=0,\eqn\lbhyp$$
similarly forms a surface which trajectories cannot cross. It partitions the
phase space into the two physically distinct regions with $\lb>0$ and $\lb<0$.\
\chapter{The $W=0$ and $Z=0$ subspaces}

If $W=0$ or $Z=0$, which corresponds physically to $\LA=0$ and $Q=0$
respectively, then one equation (either \autoe\ or \autod) drops out and the
phase space becomes 4-dimensional. In both cases, however, one direction in
the 4-dimensional subspace is ``trivial'' as a further degree of freedom can
be integrated out with a linear dependence on two of the other dimensions. In
particular, if $W=0$ then
$$V=\left(\Dm2\over\Dm3\right)X-\left(\pvy\over\pvv\right)Y+\left(\pdd\over\Dm3
\right)\cd0,\eqn\intV$$
while if $Z=0$ then
$$Y={\left(\pvy\right)\left[\Db2X-\Db3V\right]+\left[\Db3\g1-\g0\right]\cd1
\over\pyy},\eqn\intY$$
where $\cd0$ and $\cd1$ are arbitrary constants. Of course, the $W=0$ (i.e.,
$\LA=0$), system can be integrated completely [\GM]. (See also Appendix B.)
However, for our purposes here it is sufficient to stop with \intV\ and
analyse the critical points.

The only critical points at a finite distance from the origin in the full
5-dimensional phase space have both $W=0$ and $Z=0$, and so are common to
both subspaces. These critical points also have $\cP=0$, and consequently by
\lbhyp\ it follows that they are on the $\lb=0$ surface also. From \defP\ it
follows that these points are located at $X=X_0$, $Y=Y_0$, and $V=V_0$ where
$$\eqalign{&|X_0|\ge\pvvB^{1/2}|\cd0|,\cr &
Y_0=\pm\left(\pvv\over \Dm3\right)^{1/2}\left[X_0^{\ 2}-\pvvB\cd0^{\ 2}\right]^
{1/2},\cr}\eqn\XYzero$$
while $V_0$ is given by substituting \XYzero\ in \intV.

Consider the $W=0$ ($\LA=0$) subspace. If the $V$-direction is parametrised as
in \intV\ then by eliminating $V$ from the other equations we obtain an
effective 3-dimensional system:
$$\eqalignno{X'&=\Db3\Qe Z^2+X^2-\left({\Dm3\over\pvv}\right)Y^2-\pvvB
\cd0^{\ 2}&\eqname\Wa\cr
Y'&=\pvvB \Qe Z^2 &\eqname\Wb\cr
Z'&=YZ&\eqname\Wc\cr}$$
The integral curves which lie in the plane $Z=0$ are just the lines $Y=\const$
Such curves correspond physically to the spacetimes with $Q=0$ and $\LA=0$,
and the general solution for the physically interesting $\lb=1$ case was given
long ago by Buchdahl [\Bu]. The exact solutions for all values of $\lb$ are
given in [\MWa] and [\MWb]. For each value of $\cd0$ the critical points
\XYzero\ form a hyperbola in the $Z=0$ plane. From the analysis of [\MWa-\MWb]
it follows that for each value of $\cd0$ one critical point, namely the point
with\foot{The case $\g0=0$, for which the $W=0$ subspace just represents the
standard Reissner-Nordstr\"om solution can readily be treated by a separate
analysis. However, this will not concern us here.}
$$X_0=Y_0=-\pvvB{\cd0\over\g0},\qquad \g0\ne0,$$
corresponds to an horizon, $r\rarr\rH$, while the remaining critical points
correspond to a singularity at $r\rarr0$. The trajectory with an endpoint at
the horizon which lies completely in the $Z=0$ plane of course corresponds to
the Schwarzschild solution and the constant $\cd0$ is related to the
Schwarzschild radius. (See e.g.\ Fig.\ 1 in [\MWa].)

An analysis of small perturbations about the critical points $X=X_0$, $Y=Y_0$,
$Z=0$ in the 3-dimensional subspace yields the eigenvalues $0$, $2X_0$, $Y_0$.
The zero eigenvalue corresponds to the degeneracy in the $Y$ direction on the
$Z=0$ plane. Each critical point in the first and third quadrants is the
endpoint of a 2-dimensional bunch of trajectories in the 3-dimensional space,
while those in the second and fourth quadrants are saddle points with respect
to trajectories out of the $Z=0$ plane.

The critical points which lie on the sphere at infinity in the effective
3-dimensional phase space may also be found by standard means. These points
may be classified as follows:
$$\halign{$\dsp#$:&\quad $\dsp#$\hfil&$\dsp#$\hfil\cr
J\Z{1,2}&X=0,\ Y=\pm\infty,\ Z={Y\over\sqrt{\Qe \pvvB}},\qquad&\Rarr\quad
\cP={Y^2\over\pvv},\cr
K\Z{1,2}&X=\pm\infty,\ Y=X,\ Z={X\over\sqrt{\Qe \pvvB}},\qquad&\Rarr\quad
\cP={-\gg X^2\over\Db3\pvvB},\cr \noalign{\smallskip}
L\Z{1-4}&X=\pm\infty,\ Y=\pm X\sqrt{\pvv\over\Dm3},\ Z=0,\qquad&\Rarr\quad
\cP=0,\cr
M\Z{1,2}&X=\pm\infty,\ Y=0,\ Z=0,\qquad&\Rarr\quad
\cP=-X^2/\Db3,\cr}$$
The points $L\Z{1-4}$ of course correspond to the endpoints of the
one-parameter family of critical points given by \XYzero, while the points
$M\Z{1,2}$, are labelled here so as to correspond to the critical points at
infinity with the same physical properties as in [\MWa-\MWb]. Since the phase
space here describes a system which is physically different to the models
discussed in [\MWa-\MWb], the points $J\Z{1,2}$ and $K\Z{1,2}$ have no
direct correspondence to cases considered there. In Fig.~1 we plot the surface
of the sphere at infinity in terms of the coordinates $\th\Z1$ and $\ph\Z1$
defined by
$$X=\rhj\sin\th\Z1\cos\ph\Z1,\ Y=\rhj\sin\th\Z1\sin\ph\Z1,\ Z=\rhj\cos\th\Z1
,\eqn\rhmap$$
in the limit $\bar\rh\rarr1$. We will postpone the
discussion of the properties of solutions approaching these points until
the next section.

Now consider the $Z=0$ ($Q=0$) subspace. If the $Y$-direction is parametrised
as in \intV\ then by eliminating $Y$ from the other equations we obtain an
effective 3-dimensional system:
$$\eqalignno{X'&=-\LAe W^2-\Db3\cP&\eqname\Za\cr
Y'&=\left(\ggg-1\right)\LAe W^2-\Db2\cP&\eqname\Zb\cr
W'&=VW,&\eqname\Zc\cr}$$
with
$$\cP={1\over\pyy}\left[\left(\Dm1-\ggg\right)X^2-2\Db2XV+\Db3V^2+\cd1^{\
2}\right]\eqn\Zd$$
in this case.
This system, which is physically equivalent to Einstein gravity coupled to a
scalar field with a single exponential potential is of course precisely one
of the systems that have already been studied in [\MWa-\MWb], and the
properties of the solutions are identical. In addition to the critical points
common to the $W=0$ and $Z=0$ given above the following additional critical
points are found:
$$\halign{$\dsp#$:&\quad $\dsp#$\hfil&$\dsp#$\hfil\cr
L\Z{5-8}&X=\pm\infty,\ V={\Db2\pm\sqrt{\pyy}\over\Dm3},\ W=0,\qquad&\Rarr\quad
\cP=0,\cr
N\Z{1,2}&\multispan2 $\dsp X=\pm\infty,\ V=\left(\Dm1-\ggg\over\Dm2\right)
X,\ W=X\sqrt{\Dm1-\ggg\over-\LAe},$\hfil\cr \multispan2 \hbox{\hfil}&
\Rarr\quad\cP={\left(\ggg-\Db1\right)X^2\over\Db2^2}.\cr
\noalign{\smallskip}
P\Z{1,2}&X=\pm\infty,\ V=X,\ W={X\over\sqrt{-\LAe\left(\pyy\right)}},\qquad&
\Rarr\quad\cP={-\ggg X^2\over\pyy},\cr}$$
The points $L\Z{5-6}$, $N\Z{1,2}$ and $P\Z{1,2}$ have been labelled here in
precisely the same fashion as in [\MWb].
\chapter{The 5-dimensional phase space}

It is not extremely difficult to piece together the structure of the integral
curves in the full 5-dimensional phase space $\{X,Y,Z,V,W\}$ given the
existence of the various symmetries and special subspaces discussed above.
As we have already noted all critical points in the 5-dimensional phase space
which lie at a finite distance from the origin are confined to the $W=0$, $Z=0$
subspace, and this leads to great simplifications. Similarly most of the
critical points at the phase space infinity are those obtained in the previous
section; the only additional critical points are found to be the one parameter
family $L(y)$:
$$X=\pm\infty,\ Y=yX,\ V=vX,\ Z=0,\ W=0,\eqn\Lia$$
where
$$-\left[\pvv\over\Db3\right]^{1/2}\le y\le\left[\pvv\over\Db3\right]^{1/2}
\eqn\Lib$$
and\foot{Note: there is a factor $\left(\g1-\g2\right)$ missing from in front
of the square root term in the corresponding equation (2.42) in [\MWb].}
$$v={\Db2\left[\pvv\right]-\Db3\left(\pvy\right)\pm\left[\pdd\right]\sqrt{\g0
^{\ 2}+\Db3\left(1-y^2\right)}\over\Db3\left[\pvv\right]},\eqn\Lic$$
together with the isolated points
$$\halign{$\dsp#$&\quad $\dsp#$\hfil&$\dsp#$\hfil\cr
S\Z{1,2}:\ & X=\pm\infty,\ Y=X,\ V=X,&\omit\cr\omit&W=X\sqrt{\g0\over\LAe\left(
\pdd\right)},\ Z=X\sqrt{\g1\over \Qe \left(\pdd\right)},\qquad&\Rarr\quad
\cP=0,\cr \noalign{\smallskip}
T\Z{1,2}:\ &\multispan2\hbox{\hfil}$\dsp X=\pm\infty,\ Y=V=X\left(\al\Z1\over
\al\Z2\right),
W=X\sqrt{\al\Z1\over-\LAe\Db2\al\Z2},\ Z=X\sqrt{\al\Z1\al\Z3\over \Qe \al\Z2^{\
2}},$\cr \noalign{\smallskip}
\omit&\multispan2\hbox to 8true cm{\hfil} $\dsp
\qquad\Rarr\quad\cP\ =\ {-(\g0+\g1)^2\al\Z1X^2\over\al\Z2^{\ 2}}$,\cr}$$
where
$$\eqalign{\al\Z1&\equiv\Db1\gg+2\g0\g1-\Db3\ggg+\Db2^2\cr
\al\Z2&\equiv\Db2\left[\g0\gbb+\Db2\right]\cr
\al\Z3&\equiv\Db2-\g1\gbb.\cr}$$
The points $L(y)$ represent the extension of points $L\Z{1-8}$ to
the 1-parameter set of critical points which coincide with the intersection of
the $\lb=0$, $Z=0$, $W=0$ surface and the sphere at infinity.

Following [\MWa-\MWb] we will specify the asymptotic behaviour of solutions
by using the proper radius $R$ as the radial variable through coordinates
$$\dd s^2=-e^{2u}\dd t^2+e^{2v}\dd R^2+R^2g_{ij}\dd x^i\dd x^j,\eqn\coordc$$
where now $u=u(R)$ and $v=v(R)$, instead of coordinates \coorda\ or \coordb.
The properties of the solutions in the neighbourhood of the various critical
points may be readily determined. One finds three possible behaviours for the
proper radius $R$: (i) $R\rarr0$ corresponding to a central singularity as in
Table 1; (i) $R\rarr\infty$, corresponding to an asymptotic region
as in Table 2; or (ii) $R=\const$ which is true only in the case of the points
$S\Z{1,2}$. Points $S\Z{1,2}$ have $\ph={\rm constant}$ also, and they
correspond to
the endpoints of Robinson-Bertotti type solutions as is demonstrated in
Appendix A.
\midinsert \def\spil{height2pt&\omit&&\omit&&\omit&&\omit&&\omit&\cr}
\def\ar{\cr\spil\noalign{\hrule}\spil}
$$\vbox{\offinterlineskip\hrule\halign{&\vrule#&\strut\quad$\dsp#$\quad\hfil\cr
\spil&\omit&&\hbox{Values of constants}&&e^{2u}&&e^{2v}&&e^{2\ph}&\ar
&J\Z{1,2}&&\ee=+1,\ \lb=0&&R^{-2\Db3}&&R^{2\left[\G1^{\ 2}+\Db3\right]}&&R^{-
\Db2\G0}& \ar
&N\Z{1,2}&&\ggg>\Db1,\ \LAe<0, \lb=0&&R^2&&R^{2(\GGG-1)}&&R^{\Db2\G1}
&\cr \spil} \hrule}$$
\caption{Table 1}{Asymptotic form of solutions for trajectories approaching
critical points at phase space infinity from within the sphere at infinity, in
the case that $R\rarr0$.}
\endinsert
\midinsert \def\spil{height2pt&\omit&&\omit&&\omit&&\omit&&\omit&\cr}
\def\ar{\cr\spil\noalign{\hrule}\spil}
$$\hskip-12.7117pt\vbox{\offinterlineskip\hrule\halign{&\vrule#&\strut\quad
$\dsp#$\quad\hfil\cr
\spil&\omit&&\hbox{Values of constants}&&e^{2u}&&e^{2v}&&e^{2\ph}&\ar
&K\Z{1,2}&&\ee=+1,\ \lb>0&&R^{2\Db3^2/\GG}&&\const&&R^{\Db2\Db3/\G0}&\ar
&M\Z{1,2}&&\lbe>0&&\const&&\const&&\const&\ar
&N\Z{1,2}&&\ggg<\Db1,\ \LAe<0, \lb=0&&R^2&&R^{2(\GGG-1)}&&R^{\Db2\G1}&\ar
&P\Z{1,2}&&\LAe<0,\ \sign\lb=\sign\ee(\ggg-1)&&R^{2/\GGG}&&\const
&&R^{\Db2/\G1}&\ar
&T\Z{1,2}&&\LAe\al\Z1/\al\Z2<0,\ \lb=0,\ \ee\al\Z1\al\Z3>0&&R^{2\be\X1}&&R^{2
\be\X2}&&R^{\Db2^2/(\G0+\G1)} &\cr \spil} \hrule}$$
\caption{Table 2}{Asymptotic form of solutions for trajectories approaching
critical points at phase space infinity from within the sphere at infinity,
in the case that $R\rarr\infty$.\br Here
$\be\Z1=1+\Db2\left[\Db2-\g1\gbb\right]/\gbb^2$ and $\be\Z2=1-\Db2\g1/\gbb$.}
\endinsert
The only critical points which correspond to an asymptotically flat region are
$M\Z{1,2}$, and as anticipated from our earlier analysis no critical points in
Table 1 have (anti)-de Sitter asymptotics, except in the special case $\g1=0$
when the points $N\Z{1,2}$ are (anti)-de Sitter. In order to determine the
nature of the set of trajectories which have endpoints at the various critical
points it only remains to find the eigenvalues spectrum of linearised
perturbations. These results are given in Table 3.
\topinsert \def\spil{height2pt&\omit&&\omit&\cr}
\def\ar{\cr\spil\noalign{\hrule}\spil}
$$\vbox{\offinterlineskip\hrule\halign{&\vrule#&\strut\ $\dsp#$\ \hfil\cr
\spil&\omit&&\hbox to16true mm{\hfil}\hbox{Eigenvalues (with degeneracies)}&\ar
&K\Z1&&-2;\ -1,\ ({\it2});\ 1;\ {\g0\left[\g0-\Db3\g1\right]\over\Db3\left[\pvv
\right]}.&\ar &L(y)&& 0,\ ({\it2});\ 2;\ y;\ v.^*&\ar
&M\Z1&&-1,\ ({\it4});\ {1\over\Dm3}.&\ar
&N\Z1&&{-\left(\Dm1-\ggg\right)\over\Dm2}\,\ ({\it3});\ {2\over\Dm2}\left(\ggg-
1\right);\ {\g1\over\Dm2}\gbb-1.&\ar
&P\Z1&&-1,\ ({\it2});\ {-\g1\left[\pdd\right]\over\pyy};\ {-1\over2}\left[1\pm
\sqrt{9+\Db{11}\ggg\over\pyy}\,\right]\,.&\ar
&S\Z1&&-2;\ -1;\ 1; {1\over2}\left\{1\pm\sqrt{1+{8\g0\g1\gbb\over\pdd}}\,\right
\}.&\ar
&T\Z1&&-\left(\al\Z1\over\al\Z2\right), ({\it2});\ {2\over\al\Z2}\gbb\left(\pdd
\right);&\cr \spil\spil &\omit&& {\sqrt{\al\Z1}\over2\al\Z2}\left\{-\sqrt{\al\Z
1}\pm\sqrt{9\Db2\left[\gg-\ggg+\Db2\right]+\gbb^2\left(1-8\g0\g1\right)}\right
\}.&\cr\spil}\hrule}$$
\caption{Table 3}{Eigenvalues of critical points at phase space infinity. The
eigenvalues for small perturbations which are degenerate have the degeneracy
listed in brackets.\br $^*$ The values of $y$ and $v$ listed are defined by
\Lib\ and \Lic.}
\endinsert
\chapter{The structure of the phase space}

For $\LA>0$ and $\ee=1$ the only critical points at phase space infinity for
which $R\rarr\infty$ are $K\Z{1,2}$ and $M\Z{1,2}$, which both lie in the $W=0$
subspace (corresponding to $\LA=0$). The only trajectories approaching $M\Z1$
are the asymptotically flat solutions lying entirely in the $W=0$ subspace,
which of course include the familiar Gibbons-Maeda dilaton black-hole
solutions. In general, trajectories which approach the points $K\Z{1,2}$ lie
entirely within the $W=0$ subspace. However, if $\gg>\Db3\g0\g1$ then there
will be trajectories which approach this critical point from
outside the $W=0$ subspace.

On the face of it trajectories approaching the critical points $K\Z{1,2}$ are
potentially of interest, particularly for those trajectories which correspond
to the weak coupling limit $e^{2\ph}\rarr0$. Indeed, one finds that the
curvature invariants of spacetimes asymptotic to $K\Z{1,2}$ do have suggestive
properties. To be specific, in the case $D=4$, $\g0=-1$ all the components of
the Riemann tensor in an orthonormal frame go to zero as $1/R^2$ as $R\rarr
\infty$, and curvature invariants go to zero as $1/R^4$. Nevertheless, although
the $x^i=\const$ section resembles a 2-dimensional Rindler spacetime
as $R\rarr\infty$, the global structure is such that the 4-dimensional
spacetime is not asymptotically flat. The explicit solutions which approach
$K\Z1$ from within the $W=0$ subspace are found in Appendix B for arbitrary $D$
and $\g0$. It is seen that all of these correspond to naked singularities.

There are no other critical points with $R\rarr\infty$ for $\LA>0$.
All the trajectories for this system which enter the $\ee=1$ region
either end at: $M\Z{1,2}$, $K\Z{1,2}$, the Robinson-Bertotti type points
$S\Z{1,2}$, a central singularity, or at an horizon. The Robinson-Bertotti type
points exist if $$\eqalign{&\LAe/\left[\g0\left(\pdd\right)\right]>0,\cr &\ee/
\left[\g1\left(\pdd\right)\right]>0,\cr \hbox{and}\qquad&\sign\lb=\sign\ee\gbb/
\left[\g0\left(\pdd\right)\right].\cr}\eqn\spt$$ The point $S\Z1$ attracts a
3-dimensional bunch of trajectories. These are given explicitly in Appendix A.

If a trajectory has an endpoint at an horizon it will pass into the $\ee<0$
region and its asymptotics will be found by considering the $\ee<0$ critical
points. So we will now consider the various possibilities for such
trajectories. We note that it is the product of $\ee$ and $\LA$ that is
important. Thus all the remarks below about the critical points with $\ee=-1$
and $\LA>0$ also apply to the case for which $\ee=1$ and $\LA<0$.

As in the case of the models discussed in [\MWb], the structure of the phase
space is quite distinct acording whether $\ggg<(D-1)$ or $\ggg>(D-1)$.
For $\ggg<(D-1)$ the critical point $N\Z1$, lies at phase space infinity
and for small $\g1$ it attracts a 5-dimensional bunch of trajectories. For
$\g1=0$ trajectories approaching $N\Z1$ are asymptotically de Sitter.
While we cannot prove explicitly that analogues of \RN-de Sitter solutions
exist in the case $\g1=0$, the fact that the point $N\Z1$ is a 5-dimensional
attractor combined with the fact that the $Z=0$ subspace is known to contain
trajectories with two horizons which end at the point $N\Z1$ -- namely the
Schwarzschild-de Sitter solutions -- make it very likely that at least some
perturbations away from the $Z=0$ subspace end at $N\Z1$ and have two horizons.
As a check on these arguments numerical work is in progress and will be
reported elsewhere\foot{{\it Note added}: Explicit calculations
show, in fact, that in the presence of a cosmological constant {\it no}
asymptotically (anti)-de Sitter solutions exist with two horizons. In the case
of a negative cosmological constant, asymptotically anti-de Sitter black hole
solutions with a single horizon are found numerically[\JSD].} [\JSD]. In a
recent paper Okai [\Ok] reaches similar conclusions, and shows that if such
black hole solutions exist (for $\g1=0$) then there can be at most two
horizons. At this stage the reader should also recall that we are not allowing
$\g0=0$, and hence the usual \RN-de Sitter solutions are not contained in our
system. It is straightforward to repeat the entire analysis for the case $\g0=0
$ to obtain the standard results.

The points $P\Z{1,2}$ which lie in the $Z=0$ (i.e.\ $Q=0$) subspace have
asymptotics identical to those of the points $K\Z{1,2}$ if one makes the
replacement $\g0\rarr\g1/\Db3$, so the above remarks about $K\Z{1,2}$ apply
here as well. In fact, the case of the potential corresponding to a central
charge deficit in 4 dimensions, $\g1=-1$, has the precise behaviour of the
particular example discussed above. If we restrict our attention to the cases
in which $P\Z{1,2}$ can be endpoints for trajectories for physical spacetimes
with $\lb>1$ then we obtain the restriction that $\ggg\ge1$. The point $P\Z1$
is either a 3-dimensional or 4-dimensional attractor depending on the relative
signs of $\g0$ and $\g1$. For some regions of the parameter space there will be
trajectories with endpoints there for non-zero $Q$.

Finally, the asymptotics of the points $T\Z{1,2}$ are in general very
complicated. It is easy to verify that there are no particular values of $\g0$
and $\g1$ which yield either asymptotically flat or asymptotically
(anti)-de Sitter solutions. The dimension of the set of solutions with an
endpoint at $T\Z1$ varies greatly for different $\g0$ and $\g1$, but typically
it is at least 3-dimensional if $\al\Z1>0$ and $\al\Z2>0$. It is
possible to derive an exact class of solutions with endpoints at $T\Z1$ by
looking for solutions with $V=Y$ and $W=\ga Z$, where $\ga$ is constant. This
condition places a constraint on the remaining variables, and the requirement
that the field equations preserve the constraint further fixes $\ga$ to be
the same as the ratio of $W$ to $Z$ that was found for points $T\Z{1,2}$.
Since these exact solutions have $\lb=0$, however, they are of limited
physical interest, and we will not list them.

One further technical point should be made here. It was noted in section $3$
that the 5-dimensional autonomous system was not valid for $\g0=\Db3\g1$, since
the transformation between the two sets of variables becomes degenerate for
this combination of the parameters. It is possible to construct a 3-dimensional
autonomous system for this combination, and proceed to examine the phase space.
We have done this as a check and find roughly the same structure as for the
more general system that we have described here, with the exception that some
points such as $S\Z{1,2}$ are absent. Most importantly, the asymptotic forms of
the solutions given in Tables 1 and 2 are not altered.

To conclude, we have shown that, with the exception of a pure cosmological
constant, charged dilaton black holes with `reasonable' asymptotic properties,
namely an asymptocially flat or asymptotically (anti)-de Sitter behaviour, do
not exist in the presence of a Liouville-type dilaton potential. This
conclusion may seem trivial if one only considers series expansions of the type
\expand. However, it is a somewhat less trivial if one allows for the
possibility for an asymptotic behaviour of the dilaton physically consistent
with the weak coupling limit in string theory. Our conclusion is based on the
observations that (i) the critical points $M\Z{1,2}$ are endpoints only for
integral curves which correspond physically to $\V\equiv0$; and (ii) none of
the other critical points correspond to solutions with a `reasonable'
asymptotic behaviour. In the pure cosmological constant case critical points
which are asymptotically (anti)-de Sitter do exist, and given the structure of
the phase space it seems highly plausible that integral curvess with endpoints
at these critical points do include a class of charged dilaton black hole
spacetimes. However, although the method we have discussed is a useful tool for
ruling out the existence of various solutions it does not provide an obvious
way of rigourously proving the existence of solutions. Okai, who made
investigations using power series [\Ok], has also been unable to prove
unequivocally that black hole solutions do exist. Perhaps a numerical approach
is the best in such circumstances.

\bigskip\noindent{\bf Acknowledgement}: We would like to thank the Australian
Research Council and the Rothmans Foundation for financial support.

\Appendix{A}
The points $S\Z{1,2}$ are endpoints for solutions which are of
Robinson-Bertotti type. We justify this statement here by explicit derivation
of such solutions. Let us suppose that there exist solutions for which $\ph=
\hbox{constant}\ \forall\ \xi$. From \Greekcoord\ and \Latincoord\ it then
follows that
$$(D-2)X-Y-(D-3)V=0.$$
If we take the the derivative of this equation and use the field equations
\autoa--\autoe\ we obtain the condition
$$\LA W^2=\ga^2 Z^2,\qquad \ga^2=Q^2\g0/(\LA\g1).$$
Thus such solutions only exist for $\g0/(\LA\g1)>0$. Substituting back into the
field equations we find that $X=Y=V$. The solution can be put in the form
$$\eqalign{\dd s^2=&\left\{\ga^{-2\G0}\left[2\gbb Q^2\over\Db2\Db3\g1\lb\right]
^{\G0+\G1}\right\}^{1/\left[(D-3)\G1-\G0\right]}\cr &\quad\times\left\{-Z^2\dd
t^2+{\gbb\dd Z^2\over\Db3\g1\lb\left[C+\g1^{\ -1}\left(\pdd\right)Q^2Z^2\right]
}+g_{ij}\dd x^i\dd x^j\right\},\cr}\eqn\rbsoln$$
where $C$ is an arbitrary constant, and we have used the freedom of rescaling
$t$. Solutions exist only if $\gbb/(\g1\lb)>0$, or if $\lb=0$ in the case that
$\g0=-\g1$. If $\left[\pdd\right]/\g1>0$ and $\lb>0$ then the metric has the
structure of a Robinson-Bertotti spacetime, namely the product of a
2-dimensional anti-de Sitter spacetime with a $\Db2$-sphere. Similary, if
$\left[\pdd\right]/\g1<0$ then the $x^i=\hbox{constant}$ section is a
2-dimensional de Sitter space.

\Appendix{B}
Let us solve the field equations of the $W=0$ subsystem in the domain of outer
communications ($\ee=+1$). The field equations reduce to
$$\eqalignno{&\ze''=\Db3^2\lb e^{2\ze},&\eqname\jnka\cr
&\et''=2\left({\pvv}\over\Dm2\right)Q^2 e^{2\et},&\eqname\jnkb\cr
&{-\ze'^2\over\Dm3}+{\et'^2\over\pvv}+{\left(\pvv\right)\cd0^{\ 2}
\over\Dm3}+\Db3\lb e^{2\ze}-{2Q^2\over\Dm2}e^{2\et}=0,&\eqname\jnkc\cr}$$
where $\cd0$ is the integration constant defined by \intV, and $\ch$ is
completely determined in terms of $\ze$ and $\et$ by a further intergration of
\intV. Equations \jnka\ and \jnkb\ can be integrated directly with the result
$$\eqalignno{&\ze'^2=\Db3^2\left[\lb e^{2\ze}+\epk1\right],&\eqname\jnkd\cr
&\et'^2=\left(\pvv\right)\left[{2Q^2\over\Dm2}e^{2\et}+\epk2\right],&
\eqname\jnke\cr}$$
where $\ep\Z1=+1,0,-1$, $\ep\Z2=+1,0,-1$ and $k\Z1$ and $k\Z2$ are constants
which on account of the constraint \jnkc\ must satisfy the condition
$$\Db3\epk1=\epk2+\left(\pvv\over\Dm3\right)\cd0^{\ 2}.\eqn\jnkf$$
A further integration of \jnkd\ and \jnke\ yields the result
$$\lb e^{2\ze}=\cases{{k\X1^{\ 2}\over\sinh^2\left[\Db3k\X1\xixi1\right]},&
$\ep\Z1=+1$,\cr {1\over\Db3^2\xixi1^2},&$\ep\Z1=\ 0$,\cr {k\X1^{\ 2}\over\sin^2
\left[\Db3k\X1\xixi1\right]},&$\ep\Z1=-1$,\cr}$$
$${2Q^2\over\Dm2}e^{2\et}=\cases{{k\X2^{\ 2}\over\sinh^2\left[\sqrt{\pvv}\;k\X2
\xixi2\right]},&$\ep\Z1=+1$,\cr {1\over\left(\pvv\right)\xixi2^2},&$\ep\Z1=\ 0$
,\cr{k\X2^{\ 2}\over\sin^2\left[\sqrt{\pvv}\;k\X2\xixi2\right]},&$\ep\Z1=-1$.
\cr}$$
where $\xi\Z1$ and $\xi\Z2$ are arbitrary constants.

To compare with the results of Gibbons and Maeda [\GM] let us introduce a new
radial coordinate,\foot{This coordinate is denoted `$\et$' in the notation of
[\GM].}\ $\rb=\int e^{2\ze}\dd\xi$,\ so that
$$e^{2\ze}=\Db3^2\lb\left(\rb^2-\ep\Z1\rb\Z0^{\ 2}\right),\eqn\zerb$$
where $\Db3\lb\rb\Z0=k\Z1$. In the case $\ep\Z1=+1$, $\ep\Z2=+1$, the metric
functions are found to be
$$\eqalignno{&e^{2u}\propto\rrmp^{c\X0\G0\over\Db3k\X1}\muck^{-2\Db3\over\pVV},
&\eqname\yuka\cr &R^{D-3}\propto\left(\rrp\right)\rrmp^{\Db3k\X1-c\X0\G0\over2
\Db3k\X1}\muck^{\Dm3\over\pVV},&\eqname\yukb\cr
&\e^{2 \phi}=\rrmp^{-\Db2c\X0\over2\Db3k\X1}\muck^{-\Db2\G0\over\pVV},&\eqname
\yukc\cr}$$
where the constants $c$ and $A$ are given by
$$\eqalignno{&c={k\Z2\sqrt{\pvv}\over2k\Z1\Db3},&\eqname\yukd\cr&A=\exp\left[
\sqrt{\pvv}\;k\Z2\left(\xi\Z2-\xi\Z1\right)\right].&\eqname\yuke\cr}$$
One may verify that setting $k\Z2=\Db3k\Z1/\left(\pVV\right)^{1/2}$, and hence
$c={1\over2}$ and by \jnkf\ $\cd0=\Db3\g0k\Z1/\left(\pvv\right)$, one obtains
the familiar Gibbons-Maeda charged dilaton black hole [\GM].

Now consider the case $\ep\Z1=0$, $\ep\Z2=0$. One then finds that
$$\eqalignno{&e^{2u}\propto\rra^{2\Db3\over\pVV},&\eqname\zzna\cr
&R^{D-3}\propto\rb\rra^{-\Db3\over\pVV},&\eqname\zznb\cr
&e^{2\ph}\propto\rra^{\Db2\G0\over\pVV},&\eqname\zznc\cr}$$
where $\bar a=\lb\Db3^2\left(\xi\Z2-\xi\Z1\right)^2$. If $\bar a\ne0$ then
these solutions approach the points $M\Z{1,2}$ as $\rb\rarr\infty$, while if
$\bar a=0$ they approach $K\Z{1,2}$ as $\rb\rarr\infty$. It is
clear that these solutions represent naked singularites, and hence
have limited physical interest. It is straightforward to show that the
solutions obtained in the case $\ep\Z1=-1$, $\ep\Z2=-1$ similarly approach
either $M\Z{1,2}$ or $K\Z{1,2}$ as $\rb\rarr\infty$.
\refout \bigskip\bigskip \exs
\line{\fourteenrm\hfil FIGURE CAPTIONS\hfil}\medskip
Fig.\ 1.\ The hemisphere at infinity in the reduced $W=0$ ($\LA=0$)
subspace in terms of the coordinates $\th\Z1$, $\ph\Z1$ defined by \rhmap.
Although these integral curves do not correspond to physical solutions it is
nevertheless helpful to sketch them since by continuity arguments they will
determine the behaviour of the physical integral curves which lie within the
sphere at infinity but near its surface.
\end